\documentclass[11pt,preprint,flushrt]{aastex}
\usepackage{amsmath}
\usepackage{times}
\allowdisplaybreaks[1]

\newcommand\etal{{\it et al.\/}}
\newcommand{\bfg}{\mbox{\boldmath $\gamma$}}
\newcommand{\bfC}{\mbox{\bf $C$}}
\newcommand{\bfV}{\mbox{\bf $V$}}
\newcommand{\bfS}{\mbox{\boldmath $\Sigma$}}
\newcommand{\bfL}{\mbox{\boldmath $\Lambda$}}

\def\simlt{\lesssim}
\def\simgt{\gtrsim}

\begin{document}

\title{On Combining Lensing Shear Information from Multiple Filters}

\author{Mike Jarvis and Bhuvnesh Jain}
\affil{Dept. of Physics and Astronomy, University of Pennsylvania,
Philadelphia, PA 19104}
\email{mjarvis,bjain@physics.upenn.edu}

\begin{abstract}
We consider the possible gain in the measurement of lensing shear from
imaging data in multiple filters. Galaxy shapes may differ
significantly across filters, so that the same galaxy offers
multiple samples of the shear. On the other extreme, if galaxy shapes
are identical in different filters, one can combine them to improve
the signal-to-noise and thus increase the effective number density of faint, 
high redshift galaxies. We use the GOODS dataset to test these
scenarios by calculating the covariance matrix of galaxy ellipticities 
in four visual filters (B,V,i,z). We find that galaxy shapes are highly
correlated, and estimate the gain in galaxy 
number density by combining their shapes. 
\end{abstract}


\section{Introduction}

Weak gravitational lensing surveys have become a powerful tool for
testing cosmological theories.  Until recently, the majority of weak
lensing surveys have only used shape information from a single color.
Some surveys such as 
the Sloan Digital Sky Survey (SDSS)\footnote{http://www.sdss.org/}, 
the CFH Legacy survey\footnote{http://www.cfht.hawaii.edu/Science/CFHLS/}, 
COMBO-17\footnote{http://www.mpia.de/COMBO/combo\_index.html},
and COSMOS\footnote{http://cosmos.astro.caltech.edu/}
have used multiple color information for estimating photometric
redshifts \citep{Sh04,Se06,Ki07,Ma07}.  
Another use for multi-filter data has been as a cross check against
systematic effects, since many systematics are expected to be 
color dependent, whereas the lensing signature is achromatic 
\citep{Sh04,Se06}.
However, almost no study to date has made systematic use of 
the shape information from images in the different colors with the 
goal of increasing the signal-to-noise of the shear information.

Future surveys, such as 
KIDS\footnote{http://www.astro-wise.org/projects/KIDS/},
DES\footnote{http://www.darkenergysurvey.org/},
PanStarrs\footnote{http://pan-starrs.ifa.hawaii.edu/},
SNAP\footnote{http://snal.lbl.gov/} and 
LSST\footnote{http://www.lsst.org/} 
will be obtaining better imaging data in multiple filters over much
wider areas.  
Thus it behooves us to determine whether there is any significant 
advantage to using the shape information from the multiple filters
in addition to the usual magnitude information.  

There are two important regimes to consider with regards to this question.
First, with bright objects that have measurable shapes in several colors,
it might be possible to extract several independent shear estimators from
the same physical galaxy.  The plausibility of this idea comes from the
fact that galaxies look different in different colors.  For example, you
can often see starbursting regions in bluer filters which are mostly
unnoticeable in redder filters.  Similarly, giant HII regions are only
very bright in filters which includes rest frame 656 nm.  Thus a galaxy
with such regions can look quite different in this filter than it does
in other filters. 

{\bf Case A:} 
In the limiting case that galaxies look completely different in $N_{\rm f}$ 
different filters, we could obtain an independent estimate of the shear at the
location of a galaxy from every filter with a measurable shape.  
In other words, we have an independent sampling of the shape noise 
in each filter. This leads to an effective increase of the number
density of available galaxies by a factor of $N_{\rm f}$: clearly this is 
a potentially large reduction in the statistical error in shear measurement. 

{\bf Case B:} 
The other limit is that the shapes are basically the same in every color.
That is, the shape noise is sampled only once per galaxy, and we only
get one shear estimator per galaxy.  In this case we only obtain a
reduction in measurement error, which is in any case smaller than the
error due to the intrinsic ellipticities of galaxies. 
However if the galaxy shapes are the same in different filters, it is worth
considering the set of galaxies for which a good measurement of 
its shape is possible using the data in multiple filters, where no
such measurement is possible with the data in any one filter.

Since there are many more faint galaxies than bright galaxies, a large
fraction of galaxies are only slightly too faint
for a useful shape measurement.  There are obviously more 
photons in multiple filters than in just one, so it is plausible that 
we could significantly increase the number of measured shapes by using
the multi-filter data. 

In \S\ref{bright}, we investigate which of these extremes is closer 
to the truth and what the implications of this would be.
We investigate the possibile gain in effective number density of faint 
galaxies via multi-filter measurements in \S\ref{faint} and conclude
in \S4.

\section{Multi-color Shear Estimators from Bright Galaxies}
\label{bright}

Here we consider Hubble Space Telescope data from the 
GOODS\footnote{http://www.stsci.edu/science/goods/} \citep{Gi04}
survey which makes
comparably deep observations in four filters.
We used the GOODS North data below, which cover 
160 square arcminutes to depths of about $27$th magnitude 
and had about 200,000 detected galaxies.
The filters are B, V, i and z -- a fairly typical 
set of filters for an imaging survey.  The limiting magnitudes (point
source AB magnitude) in
these filters are 27.8, 27.8, 27.1 and 26.6 respectively \citep{Gi04}.

For each galaxy, we assume that 
we can measure a shape in each of these filters, which we call
$e_B^{}, e_V^{}, e_i^{}$ and $e_z^{}$, with measurement errors $\sigma_B^2,
\sigma_V^2, \sigma_i^2$ and $\sigma_z^2$ respectively.

Throughout this paper, we assume that the shape measurements have been 
converted into an unbiased shear estimator.  That is, they have been
corrected for the shear polarizability or responsivity.
That is, the expectation value of $e$ is $\gamma = \gamma_+ + i \gamma_\times$.
In this study, we used a fairly simple PSF correction, which should be 
more than sufficient for these purposes, but real applications would want 
to use a sophisticated PSF removal technique \citep[e.g.][]{Ho98, BJ02, RB03}.

If the intrinsic shapes were completely independent in different
filters, these four shape measurements could give us four independent 
estimates of the gravitational shear field at the location of the galaxy.
On the other hand, if the intrinsic shapes were identical in each filter,
then these four measurements would only give us one shear estimator with
a smaller effective measurement error.  
We use the GOODS data to see which scenario is closer to what we can expect 
for a real survey.

For a single-filter shear estimator, $\hat\gamma = e$, the estimate of
the noise on this quantity is 
\begin{equation}
{\rm Var}(\hat\gamma) = \langle |e|^2 \rangle + \sigma^2
\end{equation}

The generalization of this equation to the multi-filter case is:
\begin{align}
\hat{\bfg} &= 
\left( \begin{array}{c} e_B^{} \\ e_V^{} \\ e_i^{} \\ e_z^{} \end{array} \right) \\
{\rm Cov}(\hat{\bfg}) &= \bfC + \bfS \\
&= \left( \begin{array}{cccc} 
\langle e_B^{} e_B^* \rangle & \langle e_B^{} e_V^* \rangle &
\langle e_B^{} e_i^* \rangle & \langle e_B^{} e_z^* \rangle \\
\langle e_V^{} e_B^* \rangle & \langle e_V^{} e_V^* \rangle &
\langle e_V^{} e_i^* \rangle & \langle e_V^{} e_z^* \rangle \\
\langle e_i^{} e_B^* \rangle & \langle e_i^{} e_V^* \rangle &
\langle e_i^{} e_i^* \rangle & \langle e_i^{} e_z^* \rangle \\
\langle e_z^{} e_B^* \rangle & \langle e_z^{} e_V^* \rangle &
\langle e_z^{} e_i^* \rangle & \langle e_z^{} e_z^* \rangle \end{array} \right) +
\left( \begin{array}{cccc} 
\sigma_B^2 & 0 & 0 & 0  \\
0 & \sigma_V^2 & 0 & 0 \\
0 & 0 & \sigma_i^2 & 0 \\
0 & 0  & 0 & \sigma_z^2 \end{array} \right)
\end{align}

If the matrix $\bfC$ is close to diagonal, then we are in the first 
case described above which gives us four independent shear estimates.  
However, if the 16 values in $\bfC$ are all of similar magnitude, then we are 
in the second case, which implies that each galaxy only gives us a 
single effective shear estimate.

The GOODS data include galaxies to a limiting i magnitude of about 27, with a median 
i magnitude of 25.75.  Requiring galaxies to be observed in all four filters
lowers the median i magnitude to 25.4, not too much shallower than the
original data set.  We measured the shape in each filter and the corresponding
uncertainties using a simplified version of 
the techniques described in \citet{BJ02}. 

Using this data, we calculate the matrix $\bfC$ to be
\begin{align}
\bfC &= \left\langle \hat{\bfg} {\hat{\bfg}}^\dagger - \bfS \right\rangle \\
~&= \left( \begin{array}{cccc} 
0.25 & 0.21 & 0.20 & 0.18 \\
0.21 & 0.22 & 0.20 & 0.19 \\
0.20 & 0.20 & 0.22 & 0.18 \\
0.18 & 0.19 & 0.18 & 0.22 \end{array} \right)
\end{align}

Note that, while the shapes are intrinsically complex numbers, and therefore
$\bfC$ is as well, the imaginary components of this matrix turn out to be
insignificant (i.e. consistent with 0).  In fact, this is necessarily the
case in an isotropic universe.  Therefore we treat the matrix as
effectively real to avoid the clutter of the imaginary components.

This matrix is evidently closer to our second case,
where all 16 elements in the matrix have a similar value.
However, we can quantify this statement by diagonalizing the matrix:
\begin{align}
\bfL &= \bfV^T \bfC \bfV \\
&= \left( \begin{array}{cccc} 
0.81 & 0 & 0 & 0 \\
0 & 0.05 & 0 & 0 \\
0 & 0 & 0.03 & 0 \\
0 & 0 & 0 & 0.02 \end{array} \right)
\end{align}
This gives the noise matrix in a new basis where the measurements are uncorrelated. 
The net signal-to-noise for these four uncorrelated estimators is given by
\begin{equation}
\bfg_{\rm cosm}^T V \bfL^{-1/2} = 
\left( \begin{array}{cccc} 2.2 & 0.28 & 0.07 & 0.06 \end{array} \right)
\end{equation}
where $\bfg_{\rm cosm}^T = (~1~1~1~1~)$ is the cosmological lensing
signal in the original basis.

When combining
all of the shear data, these signal-to-noise values add in quadrature,
leading to a net increase of less than 1\% over using only the first component.
Therefore, we conclude that there is effectively no extra shear 
information available per galaxy from using the shapes measured
in multiple filters.

In general the shapes of spiral and irregular galaxies 
vary more with color than ellipticals, so perhaps this technique 
might be worthwhile for this subset of the galaxies.  To test this, we divided
the galaxy sample in half according to V-I color.  The blue half, which
should have more early type galaxies, did show
more variation than the red half, but the net signal-to-noise increase
was still less than 2\% relative to using only the first component.
So the potential improvement for the blue galaxies is still insignificant.
A similar test based on a concetration parameter was even less effective.

Note that the measurement error (in contrast to the shape error we have 
been discussing above) in the shear estimate of each galaxy
can be reduced by using the dominant eigenvalue.  The net measurement
error can be reduced by at most $\sqrt{N_{\rm f}^{}}$, where $N_{\rm f}^{}$ is the number
of filters, 4 in this case.  However, in most cases, the measurement 
errors will be different in each of the four filters, so the improvement
over the best-measured shape will be somewhat less than this.  

Furthermore, one typically cannot obtain a usable shape measurement with
measurement noise $\sigma > 0.1$.  As this is already well below the 
value of the shape noise, even reducing the measurement noise by a full factor
of two will not greatly improve the uncertainties in the final shear
statistics.

\section{Multi-color Shear Estimators from Faint Galaxies}
\label{faint}

The other case where data in multiple colors could be helpful for
weak lensing shear estimates is when the data in any one filter
is too noisy to obtain a useful shape measurement, but where the 
total signal-to-noise from all the filters would allow a useful
measurement to be obtained.  

The goal here is to increase the total number density of galaxies 
in the survey, which directly increases the signal-to-noise
of the final weak lensing data product.  
We want to know by how much we can increase the effective number density.

We will use the rule of thumb that one needs a
measurement error $\sigma < 0.1$ to obtain a useful shear estimate.  
Galaxies with a smaller signal-to-noise, which would have a measurement
error larger than 0.1 generally do not converge, or have some other 
problem which makes the estimate unreliable.
The exact value of 0.1 is unimportant for our purpose.  
Indeed, one can also increase
the number density by finding a way to increase this number while
still obtaining reliable unbiased shear estimates.  Regardless of what
number we choose, the
relative gain of using multi-filter data versus just the best filter for each
galaxy should be roughly similar.

In each filter, $k$, the galaxy has the following measured values: 
the flux, $F_k^{}$, measured in photons,
the sky noise $n_k^{}$, measuring the variance per square arcsec,
the radius, $r_k^{}$, of the best fit Gaussian, 
and the PSF radius, $r_{{\rm PSF},k}^{}$.
In terms of these values, the measurement uncertainty in the shear estimator
in a single filter is:
\begin{equation}
\sigma_k^{} = \frac{\sqrt{n_k^{}\pi r_k^2 + F_k^{}}}{F_k^{}} 
\frac{r_k^2}{r_k^2-r_{{\rm PSF},k}^2} \ .
\label{eqn:shear_error}
\end{equation}
The first factor here is simply the Poisson photon noise in the image 
divided by the flux.  For ground-based images, the noise is dominated
by the sky noise, so the second term in the numerator is negligible.
The second factor accounts for the dilution effect \citep{BJ02},
also known as smear polarizability in the KSB method \citep{KSB}. 
Basically, the PSF rounds the
image of a galaxy, so the intrinsic ellipticity estimate is always larger
than the measured observed ellipticity.  The measurement uncertainty, $\sigma$,
is also larger by the same factor.

The measurement error on a combined multi-filter shear estimate is:
\begin{equation}
\label{sig_multi}
\sigma_{\rm multi}^{} = 
\left( \sum_k^{N_{\rm f}} \frac{1}{\sigma_k^2}  \right)^{-1/2}
\ .
\end{equation}
In practice, we do not simply average the individual measurements from 
the $N_{\rm f}$ filters, as this equation might imply.  Instead, we 
perform a global maximum likelihood fit to the intrinsic shape of
the galaxy.  The observations each have a different PSF size and shape, 
so the calculations must take into account how the different PSF's 
have affected the instrinsic shape.  A more thorough description
of this process can be found in \citet{Na07}.
However, for the purposes of this paper, Equation~\ref{sig_multi}
suffices for our estimates of $\sigma_{\rm multi}^{}$.

To obtain a simplified estimate of the gain possible with
$N_{\rm f}$ filters, consider a galaxy population described by the number
magnitude relation at the flux limit
\begin{equation}
N(F) \propto F^{-2.5 s}
\end{equation}
where $F$ is the flux in a given
filter (we have dropped the subscript $k$ here). Note that $s$ is more
commonly known as the slope of the (log of) the number counts versus
limiting magnitude. 
For visible filters and galaxies at $z\simgt 1$ current data
show that $s \simlt 0.4$, giving
$N\propto F^{-1}$ or shallower. If we assume that $\sigma_k\equiv \sigma$ 
is the same in all filters, we get $\sigma_{\rm multi}^2 = 
\sigma^2/N_{\rm f} \sim 1/(F^2 N_{\rm f})$ (approximating from equation
\ref{eqn:shear_error}).  
Thus using $N_{\rm f}$ filters lowers the effective flux limit to
$F/\sqrt{N_{\rm f}}$, which corresponds to an increase in number density
of $n\rightarrow n N_{\rm f}^{2.5 s/2}$. For $s\simeq 0.4$, the increase in
$n$ is a factor of $\sqrt{N_{\rm f}}$; it is lower for smaller $s$. 
Thus a rough guide to the maximum 
gain possible from multiple filters is that by using four filters with 
comparable signal-to-noise shape measurements, 
the effective number density of galaxies doubles. 

For the GOODS data we estimate how much the number density 
can be increased by using the galaxies which have 
$\sigma_{\rm multi}^{} < 0.1$, 
but for which all $\sigma_k^{} > 0.1$.
We also impose an additional restriction, which we have not discussed
in this paper, but which experience has shown is required for unbiased
shear estimates.  Namely, we only use galaxies which have 
$r > 1.2 r_{\rm PSF}$.
When this criterion gives different answers in different filters, we 
do consider the data from those filters which pass the test  (i.e., we
do not reject a galaxy entirely just because the PSF is too large 
in one of the filters.)

Using the GOODS data, we find that the number density increases by 30\%
if one uses the multi-filter shear estimates rather than only those with
a measurable shear in at least one filter. 
For this data set, most of the galaxies with 
measurable shapes in any filter were measureable in the z filter. 
In fact if the multi-filter shapes are compared to the shape measured
just in the z filter the net increase in number density is 34\%, only
slightly higher (it would of course be significantly higher if the
comparison was made with any of the other filters).  
This increase however is signficantly smaller than the idealized
estimate of a factor of two above, because the measurement error
is far from constant in the different filters. This is mostly due to
the galaxy spectrum being far from flat; the z filter was the
shallowest of the four in terms of AB magnitude, but it still
typically had the highest signal to noise. For shallower surveys,
such as planned near-term surveys from the ground, the median
redshifts will be lower. Hence one would expect that the signal to
noise will peak at lower wavelengths, but the effective number of
filters will probably still be at most 2. 

The improvement is likely to be greater for surveys with 
more than four, presumably narrower, 
filters and for specific galaxy populations. It does not appear to be
much larger based on
tests with simulated images kindly provided by D. Johnston which used 
realistic shapelet models for the galaxy shapes based on UDF images.
For these data, we found that the gain in galaxy number density 
with six narrow filters was still about 30\%.  

A second factor that
can provide benefits is that 
the fainter galaxies are at higher redshift than the median for the
whole sample. Hence they carry greater lensing signal and offer a
wider redshift range for tomography.  Therefore the increase in 
number density may underestimate the resulting increase in signal-to-noise
for cosmological parameter estimation.

We also note that we assumed that the 
lower limit in $\sigma$ from the multi-filter data is the same as for
single filter data.  This sounds like an obvious assumption, but in fact
it is not yet true.  Current algorithms for measuring a shape from 
multiple images (whether the same or different filters) have more trouble
converging than algorithms which only use data from a single image (keeping
total signal-to-noise constant).  We expect that it should be possible to 
improve these algorithms to where the multi-image algorithm is as effective
as the single-image algorithm, but it does require further work.

\section{Discussion}

We found based on the sample of galaxies from GOODS that 
galaxy shapes are highly correlated between visual
filters. Hence there is no significant advantage to having 
multi-filter shape information for bright galaxies in a weak 
lensing survey.  However, for the fainter galaxies that are close to
the signal-to-noise threshold, we find that there is a significant 
gain in using multiple filters because one can measure shapes for
galaxies below the threshold in any one filter.  There are
34\% more galaxies with usable shear estimates in the GOODS B,V,i,z data 
by combining shape information from all four filters
than by simply using shapes measured in z, the single best filter 
for all galaxies.  

The specific results are somewhat tied to the properties
of the GOODS data set.  For surveys which have more filters or 
which have a different magnitude limit or noise properties, the
exact benefit may vary.  For ground based surveys the
sky noise and seeing will play a role as well. However, based in part on
tests with simulated images, we believe 
that the basic result will hold -- there are improvements 
to be had due to an increase in galaxy number density 
at the faint end, but they are below the factor of two (or
more) one might have expected in an idealized case. The 
galaxy properties underlying this conclusion are: 
(i) galaxy shapes are highly correlated
across the visual filters, but, (ii) the signal-to-noise for shape
measurements varies so that only a few filters contribute
significantly for a given galaxy. 

There is an additional gain in lensing measurements because the
faint galaxies recovered using multi-filter shapes 
are at higher redshifts than the median redshift of
the whole sample. These galaxies carry greater lensing signal, and
more usefully, enable more tomographic bins to be used for
cosmological analysis. Surveys of modest depth, for which these faint
galaxies lie at redshifts $\simgt 1$, will have the greatest improvment
in accuracy of parameters such as the dark energy equation of state. 
The key factors that impact the gain for a particular survey are: the redshift
distribution of the galaxy sample, the typical galaxy sizes relative
to the seeing at the faint end, and the signal-to-noise per filter needed for
the desired accuracy in photo-z's (the photo-z measurement is not
considered in this study).  


Finally, we emphasize that systematic errors can be checked and partially
eliminated with multi-filter data. The SDSS analyses of galaxy-galaxy
lensing (Fisher et al. 2000) used the consistency of the tangential
shear profile in the SDSS g', r' and i' filters as one of the main
checks of systematics. In future surveys that aim to measure shears
with percent level accuracy, comparison of shape measurements will 
be valuable in checking for the effect of color dependent PSFs --
these affect stars and galaxies differently and are not well
quantified at the present time. 

\acknowledgements
We thank Gary Bernstein, Sarah Bridle, 
Dave Johnston, Jason Rhodes, Fritz Stabenau, 
Masahiro Takada, Ludo van Waerbeke and David Wittman for helpful discussions. 
We also thank the anonymous referee for very useful feedback.
This work is supported in part by NSF grant AST-0607667, the 
Department of Energy and the Research Corporation.


\newpage
\begin{thebibliography}{}

\bibitem[Bernstein \& Jarvis(2002)]{BJ02} Bernstein, G.~M., 
\& Jarvis, M.\ 2002, \aj, 123, 583 

\bibitem[Fischer et al.(2000)]{Fi00} Fischer, P. et 
al.\ 2000, \aj, 120, 1198

\bibitem[Giavalisco et al.(2004)]{Gi04} Giavalisco, M., et 
al.\ 2004, \apjl, 600, L93 

\bibitem[Hoekstra \etal(1998)]{Ho98}
Hoekstra, H., Franx, M., Kuijken, K., \& Squires, G. 1998, \apj, 504, 636

\bibitem[Kaiser, Squires, \& Broadhurst(1995)]{KSB}
Kaiser, N., Squires, G., \& Broadhurst, T. 1995, \apj, 449, 460

\bibitem[Kitching et al.(2007)]{Ki07} Kitching, T.~D., 
Heavens, A.~F., Taylor, A.~N., Brown, M.~L., Meisenheimer, K., Wolf, C., 
Gray, M.~E., \& Bacon, D.~J.\ 2007, \mnras, 376, 771 

\bibitem[Massey et al.(2007)]{Ma07} Massey, R., et al.\ 
2007, \apjs, 172, 239

\bibitem[Nakajima \& Bernstein(2007)]{Na07} Nakajima, R., \& 
Bernstein, G.\ 2007, \aj, 133, 1763 

\bibitem[Refregier \& Bacon(2003)]{RB03}
Refregier, A. \& Bacon, D. 2003, \mnras, 338, 48

\bibitem[Semboloni et al.(2006)]{Se06} Semboloni, E., et 
al.\ 2006, \aap, 452, 51 

\bibitem[Sheldon et al.(2004)]{Sh04} Sheldon, E.~S., et al.\ 
2004, \aj, 127, 2544 

\end{thebibliography}
\end{document}